\documentclass[secnumarabic,aps]{revtex4}

\usepackage[tbtags]{amsmath}
\usepackage{amssymb,bm}
\usepackage{graphics}
\usepackage{color}

\newcommand{\mean}[1]{\left\langle#1\right\rangle}

\newcommand{\dd}{\mathrm{d}}
\newcommand{\br}{\bm{r}}
\renewcommand{\bf}{\bm{f}}
\newcommand{\HH}{\hat{\mathcal H}}
\newcommand{\aA}{{\mathcal A}}
\newcommand{\FF}{{\mathcal F}}
\newcommand{\ds}{{d_{s}}}
\newcommand{\dsd}{\frac{d_{s}}{2}}

\newcommand{\ad}{a^{\dagger}}
\newcommand{\bd}{b^{\dagger}}
\newcommand{\bra}[1]{\left\langle {#1} \right| }
\newcommand{\ket}[1]{\left| {#1} \right\rangle }
\newcommand{\btf}{\tilde{\bm{f}}}

\newcommand{\beq}{\begin{equation}}
\newcommand{\eeq}{\end{equation}}
\newcommand{\beqn}{\begin{eqnarray}}
\newcommand{\eeqn}{\end{eqnarray}}

\begin{document}

\title{Fast algorithms for classical $X\to 0$ diffusion-reaction processes}

\author{Fabrice Thalmann$^{(1)}$ and Nam-Kyung Lee$^{(2)}$}
\affiliation{
$^{(1)}$ Universit{\'e} Strasbourg 1, Institut Charles Sadron, CNRS UPR 22,
23 rue du Loess, BP~84047, F-67034 Strasbourg Cedex, France\\
$^{(2)}$ Department of Physics, Institute of Fundamental Physics, Sejong
University 143-747, Seoul, South-Korea}
\date{\today}
\begin{abstract}
  The Doi formalism treats a reaction-diffusion process as a quantum many-body
  problem. We use this second quantized formulation as a starting point to
  derive a numerical scheme for simulating $X\to 0$ reaction-diffusion
  processes, following a well-established time discretization procedure. In
  the case of a reaction zone localized in the configuration space, this
  formulation provides also a systematic way of designing an optimized,
  multiple time step algorithm, spending most of the computation time to
  sample the configurations where the reaction is likely to occur.
\end{abstract}
\maketitle

\section{Introduction}
\label{sec:Introduction}

The reaction diffusion process $X\to 0$ with space dependent reaction rate,
models a number of interesting chemical and biological situations.  It models
for instance the capture dynamics of a mobile ligand diffusing in a
configuration space until being captured by a static receptor. This
configuration space may not just be the three dimensional volume containing
the molecules which participate to the reaction, but must be considered in a
wider perspective.  For example, in the case of a ligand bound to one end of a
polymer spacer while the other end is grafted onto a surface (bridging
situation~\cite{2001_Jeppesen_Marques}), the diffusive dynamics naturally
takes place in a configuration space taking into account all the internal
degrees of freedom of the chain, even though the reaction rate mainly depends
on the position of the polymer end bearing the
ligand~\cite{2004_Moreira_Marques}. Biology and bio-technologies provide
numerous examples of adhesion mechanisms with kinetics limited by similar
reaction-diffusion dynamics~\cite{1958_Alberty_Hammes, 1982_Chou_Zhou}.  The
most standard immunodiagnosis assays utilize the aggregation processes based
on antigen-antibody specific adhesion. It is now possible to set up finely
tuned experiments with a high enough temporal resolution for the comparison of
reaction kinetics with analytical or numerical
predictions~\cite{1998_Chesla_Zhu,2008_Lee_Marques}.

Polymer chemistry also offers a wide set of situations which can be
advantageously modeled by a $X\to 0$ process. The cyclization reactions
occurring during macromolecules synthesis are the
examples~\cite{1974_Wilemski_Fixman:1,1974_Wilemski_Fixman:2}. The situation
can be further complicated if an external hydrodynamic flow is present, or
more generally, if conservative or dissipative force fields are known to act
upon the system under study~\cite{1996_Frederickson_Leibler}, then calling for
intensive numerical
simulations~\cite{1984_Northrup_McCammon,1996_Pastor_Szabo}.

We present in this paper a general method for finding efficient numerical
simulation algorithms, using a non hermitic Hamiltonian formulation of the
stochastic process introduced by Doi~\cite{1976_Doi:1}, and based on a Trotter
splitting of the discrete time evolution operator. This strategy has proved
successful when applied to Brownian and DPD (dissipative particle dynamics)
dynamics, leading to the  generation of algorithms, with controlled
time-reversal or accuracy properties~\cite{2006_DeFabritiis_Coveney,
  book_Frenkel_Smit,2003_Ricci_Ciccotti,2007_Thalmann_Farago}.

Moreover, we suggest for this problem an original multiple time step
algorithm, based on some internal state switching $A \leftrightarrow B$, which
comes as a straightforward extension of the physical reaction-diffusion
process. We then use a different integration time step, depending on the
internal variable state, either fast or slow.

The essential strategy is to introduce fast coarse grained sampling away from
the reaction zone (while keeping slow sampling near the reaction zone) in
order to achieve the desired computational efficiency and accuracy.  This
algorithm comes as a new member in a family of optimized multiple time step
algorithms, such as the RESPA~\cite{1990_Tuckerman_Berne,
  1992_Tuckerman_Martyna}, and previously restricted to a fixed number of
moving particles.

In this work, we restrict our approach to the case of $X\to 0$ non-interacting
particles.  However it can be easily extended to more general cases without
difficulty.  Although the formalism of Doi is not new, it is, to our
knowledge, the first time that it serves as a starting point for designing a
specific integration scheme for reaction-diffusion simulations.

We introduce in Section~2 the Doi formalism~\cite{1976_Doi:1,1976_Doi:2} and
derive the corresponding single time step algorithm. We underline in Section~3
the danger of naive incorporation of multiple integration time-steps for
different regions of the configuration space, and show how to properly
formulate multiple time step algorithms.

Section~4 demonstrates the validity of the whole scheme for particles
diffusing on a spherical surface.  The simulation time can be reduced by a
large factor without noticeable artifact by choosing proper step sizes. The
estimate of the expected speeding up efficiency is provided.  Perspectives are
outlined in our conclusive Section.


\section{Formalism}
\label{sec:Formalism}

The standard master equation for the $X\to 0$  process is
\beq
\partial_t n(\br,t) = D\nabla^2 n(\br,t) -\mu\nabla (n(\br,t)\bf(\br,t))
-S(\br) n(\br,t),
\label{eq:standardMasterEquation}
\eeq
where $n(\br,t)$ is the probability of finding a particle at a position $\br$,
at time $t$, $S(\br)$ is the space dependent reaction rate, $D$ the diffusion
constant, $\mu$ the mobility coefficient, linked to $D$ by the Einstein
relation $D=k_B T\mu$, and $\bf(\br)$
a space dependent, constant force field.
Eq.~(\ref{eq:standardMasterEquation}) is a Fokker-Planck-Smoluchovski
diffusion process, with an additional sink term accounting for particle
decay. Usually, the rate $S(\br)$ is a sharp function of $\br$, and
represents, for instance, the very localized region in configuration space
where either a binding or a chemical reaction can take place, with a precision
lying in the range of the angstrom.  The diffusive behavior underlying
eq.~(\ref{eq:standardMasterEquation}) corresponds to an overdamped Brownian
diffusion, in which the inertia of the diffusive particle plays no part.

The $X\to 0$ process does not conserve the number of particles, but is not a
many-body problem, as there is no interaction affecting either the spatial
motion or the reaction rate. Thus, following the idea of Doi, we introduce the
second-quantized time evolution operator, that we subsequently call
``Hamiltonian'' by reference to standard quantum
mechanics~\cite{1976_Doi:1,1976_Doi:2,1996_Cardy} (it is also referred as a
Liouvillian in related ref.~\cite{1985_Peliti}).
\beq
\HH = \int \dd\br \bigg[D (\nabla\ad(\br)\cdot\nabla a(\br))
-\mu (\bf(\br)\cdot\nabla\ad(\br))a(\br)
+S(\br)(\ad(\br)a(\br) -a(\br))\bigg].
\label{eq:DoiPelitiProcess}
\eeq

In eq.~(\ref{eq:DoiPelitiProcess}), the first term corresponds to the Brownian
diffusion term $(\HH_D)$, the second one to the force induced drift term
$(\HH_F)$, and the last one is the space dependent reaction $X\to 0$
($\HH_R$). The operators $a(\br)$ and $\ad(\br)$ are bosonic creation and
annihilation operators.  Here, the use of bosonic operators means that
multiple occupancy at a given position $\br$ is possible. However, in the
absence of reverse reaction $0\to X$, or any other reaction creating
particles~X, the total number of particles present in our system can only
decrease. If one starts at $t=0$ with a single particle at the origin
$\br=\br_0$
(\textit{i.e.} $n(\br,\br_0)=\delta(\br-\br_0)$).
\beq
\ket{\Phi(0)} = \ad(\br_0)\ket{0},
\eeq
then the evolution operator $\exp(-\HH t)$ of our system, either moves or
destroys this particle, and double occupancy never happens. Thus,
for a single particle reaction dynamics, 
formally it makes no difference here whether one chooses to use bosons or
fermions in eq.~(\ref{eq:DoiPelitiProcess}).

The connection between the second-quantized formalism and the standard
formalism is made with
\beq
n(\br,t) = \bra{0}\exp\bigg(\int\dd\br'a(\br')\bigg)a(\br) \ket{\Phi(t)},
\eeq
where $\ket{\Phi(t)} = \exp(-\HH t)\ket{\Phi(0)}$ represents the state of the
system at time $t$~\cite{1996_Cardy}. As double occupancy is precluded, one can
safely replace the above expression by
\beq
n(\br,t) = \bra{0}a(\br)\ket{\Phi(t)} = \bra{0}a(\br) e^{-\HH\, t}
\ket{\Phi(0)}.
\label{eq:densitySingleOccupancy}
\eeq
and define a propagator, for positive values of~$t$
\beq
G(\br_2,\br_1;t) = \bra{0}a(\br_2)e^{-\HH\, t}\ad(\br_1)\ket{0}.
\label{eq:Propagator}
\eeq

In eq.~(\ref{eq:DoiPelitiProcess}), the Hamiltonian is a sum of three
non-commuting terms $\HH_D$, $\HH_F$ and $\HH_R$, and $\exp(-\HH t)$ is not
known analytically. Following the Trotter-Strang idea, we split the
evolution operators as
\begin{multline}
e^{-(\HH_D+\HH_F+\HH_R)\, t} =
\bigg[e^{-(\HH_D+\HH_F+\HH_R) \Delta t}\bigg]^{t/\Delta t}\\
\simeq \bigg[ e^{-\frac{\Delta t}{2}\HH_R}
 \cdot e^{-\frac{\Delta t}{2}\HH_F}
 \cdot e^{-\Delta t\HH_D}
 \cdot e^{-\frac{\Delta t}{2}\HH_F}
 \cdot e^{-\frac{\Delta t}{2}\HH_R} \bigg]^{t/\Delta t}.
\label{eq:splittingRFDDFR}
\end{multline}

The approximation becomes exact for $\Delta t\to 0$, and the above splitting
is known to be exact up to terms of order $\Delta t^3$.
Expression~(\ref{eq:splittingRFDDFR}) has a clear algorithmic interpretation.
It tells us to iterate a large number of times ($N=t/\Delta t$) a sequence
of five steps, which altogether are supposed to reproduce faithfully the
dynamics of the process over a time interval $\Delta t$. The error committed
when doing this approximation is of order $\Delta t^3$ and originates from the
fact that we have split the exponential of a sum of three non-commuting terms.

We now consider each one of these steps: $\exp(-\Delta t \HH_R/2)$ represents
the action of the reaction term, at a fixed position, during a time interval
$\Delta t/2$. Then, the survival probability of a particle initially at $\br$
is $\exp(-S(\br)\Delta t/2)$, and the probability of having reacted is just
\hbox{$1-\exp(-S(\br)\Delta t/2)$}.

The term $\exp(-\Delta t \HH_F/2)$ represents a pure drift along the (constant
in time) force field $\bf(\br) $. This is tantamount to solving the first
order differential equation $\dot{\br}(t')= \mu\bf(\br(t'))$, with initial
condition $\br(t'=0) = \br$, during a time interval $\Delta t/2$. We can check
that, at the desired order of accuracy (\textit{i.e.} $\Delta t^2$), this
corresponds to a shift $\br \to \br +\mu\Delta t \btf(\br)/2$, where the
pseudo force field $\btf(\br)$ is closely related to the original force field
$\bf(\br)$, according to the Heun, or Collatz scheme (see Appendix~II and
ref.~\cite{book_Stoer_Bulirsch}).

The term $\exp(-\Delta t \HH_D)$ represents a pure Brownian diffusion in flat
space, which is nothing but a Gaussian (Wiener) process with variance
$2D\Delta t$.

The splitting in Eq.~(\ref{eq:splittingRFDDFR}) can be cast in the algorithm
presented in Table~\ref{table:algo-I}. At the end of this procedure, the
position of the particle is
\beq
\br(t+\Delta t) = \br(t) + \frac{\mu\Delta t}{2}\bigg[\btf(\br(t))+
\btf(\br(t) +\mu\btf(\br(t))\Delta t/2 +\Delta \br^D)\bigg] + \Delta \br^D,
\eeq
and the survival probability is $\exp(-[S(\br(t))+S(\br(t+\Delta
t))]\Delta t/2)$. If a reaction takes place, the reaction time must be set to
\beq
t_R=t+\frac{\Delta t}{2},
\eeq
and the reaction diffusion process stops. In fact, we cannot define the
reaction time more accurately than that, because any attempt to link the
reaction time to one particular step in the previous algorithm would be
pointless.  The five steps of Algorithm~I are the
consequence of a formal, mathematical splitting of~(\ref{eq:splittingRFDDFR}),
and does not represent a consecutive sequence of events in real time.

Such an algorithm should prove successful when applied to smooth,
differentiable space varying functions $S(\br)$ and $\bf(\br)$.
The accuracy of its predictions is expected, when averaged over the different
Brownian paths and random reaction times $t_R$, to reproduce the results of
the exact, continuous time process, up to and including the order $\Delta
t^2$. These algorithms are known, in applied mathematics, as ``weak order two
algorithms''~\cite{book_Kloeden_Platen}.


\section{Optimization and multiple time step algorithms}
\label{sec:Optimization}

The accuracy of the above numerical scheme is limited by the regularity of the
functions $S(\br)$ and $\bf(\br)$ appearing in the problem.  The
characteristic length scales $\Lambda_S$ or $\Lambda_{\bf}$ upon which these
functions vary set the upper bound for the discrete spatial increment $\Delta
\br$, which in turn set the upper bound for the discrete time step $\Delta t$
of our dynamics.

In capture problems, it commonly happens that the reaction zone occupies only
a tiny fraction of the configuration space, with a length scale $\Lambda_S$
of the size of the angstrom.

\begin{figure}
\begin{center}
\resizebox{\textwidth}{!}{\includegraphics*{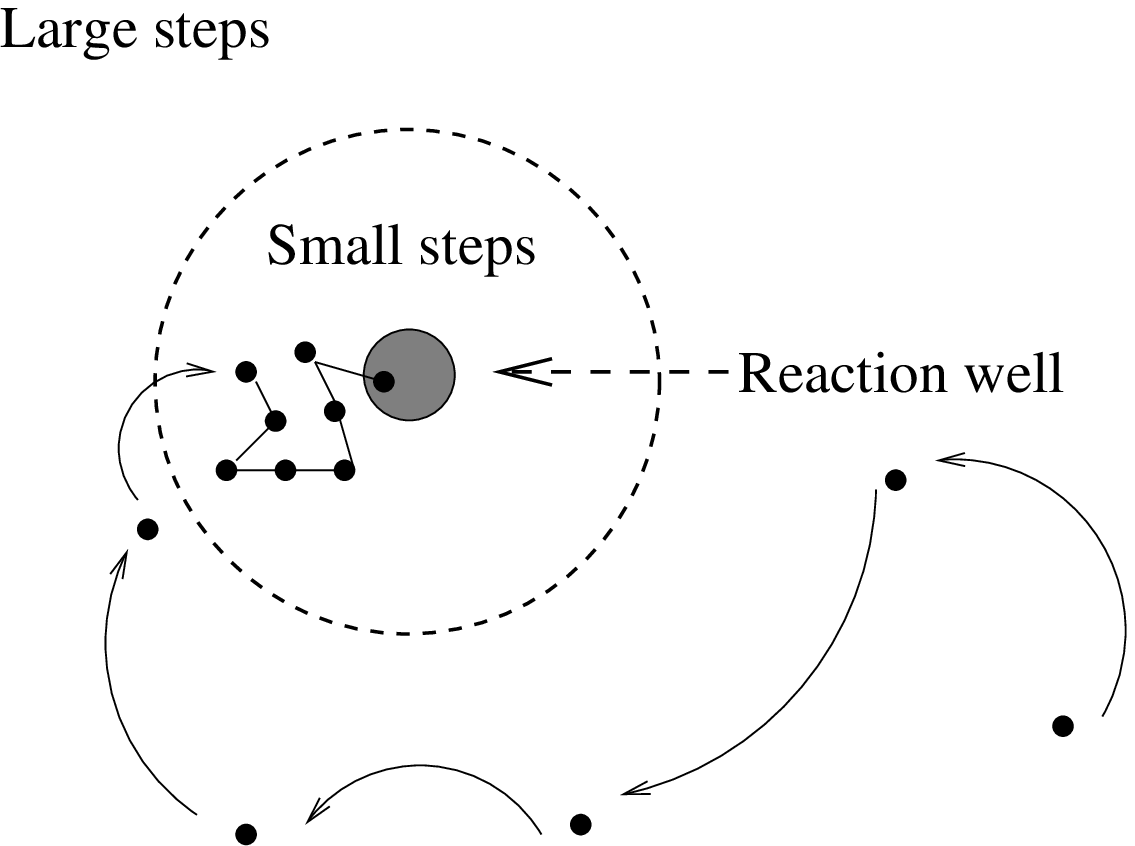}}
\caption{}
\label{fig:multistep}
\end{center}
\end{figure}

Using a large time step $\Delta t$ enables a fast sampling of the
configuration space. However the risk of missing the reaction zone increases
accordingly overlooking the possibility for the Brownian path, which is a
continuous process, to come across the reaction zone and react
(Fig.~\ref{fig:multistep}).

In practice, the time step $\Delta t$ must be chosen in order to
keep the typical length step $||\br(t+\Delta t)-\br(t)||$ smaller than
$\Lambda_S$ by a factor two or three. On the other hand, the choice of a small
time step can make the computational effort unbearable if $L\gg \Lambda_S$, and
when nothing much happens outside the reaction zone. For those reasons, it is
desirable to have a multiple time step algorithm at our disposal.

A naive attempt to make a multiple time step algorithm is to divide the
configuration space into two zones: a slow zone where one applies the
algorithm with a small time step $\Delta t$, and a fast zone where one applies
the algorithm with a larger time step $\FF \Delta t$, where $\FF$ is an
arbitrary acceleration factor. Convenient choices for $\FF$ are powers of $2$,
such as $\FF=8,16\ldots$, to enforce commensurability between small and fast
dynamics.

As a test case, we apply this idea to a one dimensional particle diffusing in
the interval $x\in[-0.5,0.5]$ (periodic boundary conditions) with ``slow'' and
``fast'' domains and an acceleration factor $\FF=8$. The large time step
applies for a particle in $|x| > 0.2$ and the small step for particles in
$|x|<0.2$, which would correspond to a reaction zone close to $x=0$.
Figure~\ref{fig:sharp} shows the result of this simulation. A sharp
discontinuity between fast and slow zones leads to an inhomogeneous density of
particles. This artifact is located near the slow/fast boundary, with width
$\sim\sqrt{2 \FF D \Delta t}$ given by a diffusion length, and due to a local
breakdown of the detailed balance relations, where incoming particles large
moves are not balanced by the reverse small moves.

In order to suppress this artifact, we propose to introduce a region, called
``exchange zone'', of size $\Lambda_w$ where fast and slow dynamics overlap.
This is achieved by adding to the diffusing particles an internal state
variable, $\sigma=A$ or $B$, switching randomly between a state $A$ and a
state $B$, like in a reversible chemical reaction $A\leftrightarrow B$.

The standard master equation of the process becomes:
\beqn
\partial_t n_A(\br,t) &=& D\nabla^2  n_A(\br,t) -\mu\nabla(\bf(\br) n_A(\br,t))
-S(\br)  n_A(\br,t)) \notag\\
& & + w_{AB}(\br)n_B(\br,t) -w_{BA}(\br)n_A(\br,t);\notag\\
\partial_t n_B(\br,t) &=& D\nabla^2  n_B(\br,t) -\mu\nabla(\bf(\br) n_B(\br,t))
-S(\br)  n_B(\br,t)) \notag\\
& & -w_{AB}(\br)n_B(\br,t) +w_{BA}(\br)n_A(\br,t).
\label{eq:twoStatesMasterEquation}
\eeqn

The internal state ($\sigma=A$ or $\sigma=B$) does not alter the diffusion nor
the reaction properties of the particle. $\sigma=A$ or $B$ is a hidden
variable, while all the space dependent properties remain unchanged compared
with~eq.~(\ref{eq:standardMasterEquation}). We loosely call particle~A a
particle with $\sigma=A$ and particle~B a particle with $\sigma=B$.  In the
slow diffusion limit, the equilibrium molar fraction of particles $A$ and $B$
is fixed by the exchange rates $w_{AB}(\br)$ and $w_{BA}(\br)$, and given by
\beqn
x_A = \frac{n_A}{n_A+n_B} &=&
\frac{w_{AB}(\br)}{w_{AB}(\br)+w_{BA}(\br)};
\notag\\
x_B = \frac{n_B}{n_A+n_B} &=&
\frac{w_{BA}(\br)}{w_{AB}(\br)+w_{BA}(\br)}.
\label{eq:molarFractions}
\eeqn
The exchange rates $w_{AB}(\br)$ and $w_{BA}(\br)$ can be arbitrarily chosen,
and tailored to compromise between an strong conversion rate ($w$ must be
large), and a relative smooth spatial variation ($w$ cannot vary too sharply
with $\br$). We keep the size of the exchange zone $\Lambda_w$ two to three
times larger than the spatial increment of the particles to preserve the total
particle number.

The  second-quantized Hamiltonian corresponding to this situation is:
\beq
\HH_{\sigma} = \HH_{A,R}+\HH_{A,D}+\HH_{A,F}+\HH_{A,Ex}
+ \HH_{B,R}+\HH_{B,D}+\HH_{B,F}+\HH_{B,Ex},
\label{eq:switchingHamiltonian}
\eeq
with components
\beqn
\HH_{A,R} &=& \int\dd\br\bigg[S(\br)(\ad(\br)a(\br) -a(\br))\bigg];\\
\HH_{A,D} &=& \int\dd\br\bigg[D (\nabla\ad(\br)\cdot\nabla a(\br))\bigg];\\
\HH_{A,F} &=& \int\dd\br\bigg[-\mu(\bf(\br)\cdot\nabla\ad(\br))a(\br)\bigg];\\
\HH_{A,Ex} &=& \int\dd\br\bigg[w_{BA}(\br)\bigg(\ad(\br) a(\br) -\bd(\br)
a(\br)\bigg)\bigg];\\
\HH_{B,R} &=& \int\dd\br\bigg[S(\br)(\bd(\br)b(\br) -b(\br))\bigg];\\
\HH_{B,D} &=& \int\dd\br\bigg[D (\nabla\bd(\br)\cdot\nabla b(\br))\bigg];\\
\HH_{B,F} &=& \int\dd\br\bigg[-\mu(\bf(\br)\cdot\nabla\bd(\br))b(\br)\bigg];\\
\HH_{B,Ex} &=& \int\dd\br\bigg[w_{AB}(\br)\bigg(\bd(\br) b(\br) -\ad(\br)
b(\br)\bigg)\bigg].
\eeqn
where $b,\bd$ are bosonic creation/annihilation operators associated with
particle~B, and $a,\ad$ associated with particle~A.  Up to this point,
particles $A$ and $B$ obey the same dynamics, and one is not faster than the
other.  The numerical scheme originates from the splitting of
$\exp(-\FF\HH_{\sigma}\Delta t)$ as follows.
\begin{multline}
\exp(-\FF\HH_{\sigma}\Delta t) =\\
\bigg\lbrace%
e^{-\Delta t\HH_{B,Ex}/2}%
\bigg[%
e^{-\Delta t\HH_{B,R}/2}%
e^{-\Delta t\HH_{B,F}/2}%
e^{-\Delta t\HH_{B,D}}%
e^{-\Delta t\HH_{B,F}/2}%
e^{-\Delta t\HH_{B,R}/2}%
\bigg]%
e^{-\Delta t\HH_{B,Ex}/2}%
\bigg\rbrace^{\FF/2}\\%
\times%
e^{-\FF\Delta t\HH_{A,Ex}/2}%
\bigg[%
e^{-\FF\Delta t\HH_{A,R}/2}%
e^{-\FF\Delta t\HH_{A,F}/2}%
e^{-\FF\Delta t\HH_{A,D}}%
e^{-\FF\Delta t\HH_{A,F}/2}%
e^{-\FF\Delta t\HH_{A,R}/2}%
\bigg]%
e^{-\FF\Delta t\HH_{A,Ex}/2}\\%
\times\bigg\lbrace%
e^{-\Delta t\HH_{B,Ex}/2}%
\bigg[%
e^{-\Delta t\HH_{B,R}/2}%
e^{-\Delta t\HH_{B,F}/2}%
e^{-\Delta t\HH_{B,D}}%
e^{-\Delta t\HH_{B,F}/2}%
e^{-\Delta t\HH_{B,R}/2}%
\bigg]%
e^{-\Delta t\HH_{B,Ex}/2}%
\bigg\rbrace^{\FF/2}.%
\label{eq:splittingMultistep}
\end{multline}
The above splitting has a straightforward algorithmic interpretation. These
terms indexed by $B$ only acts on the particles whose internal state is $B$
and leave the $A$ particles unchanged with both position and internal states.
The converse is true.

One recognizes three occurrences of the splitting~(\ref{eq:splittingRFDDFR}),
enclosed in brackets $[\cdot]$, which do not depend on the internal state of
the particles, but only on the time interval (respectively $\Delta t$ or
$\FF\Delta t$).  The splitting procedure has already been detailed in
Section~\ref{sec:Formalism}.  Therefore, any implementation of the single time
step algorithm can be reused in the multiple time step situation, while
providing also a natural checking procedure for this more complex algorithm.

The effect of $\exp(-\Delta t\HH_{B,Ex})$ is to exchange the internal state of
$B$ particles to $A$, with a space dependent probability equal to
$1-\exp(-\Delta t w_{AB}(\br))$, corresponding to a waiting time $\Delta t$.
Meanwhile, particles $A$ are not affected and stay at the same position $\br$.
The resulting algorithm is summed up in Table~\ref{table:algo-II}.

The dynamics of particles $B$ is treated with a discrete time step $\Delta t$,
while the dynamics of particles $A$ is treated with a larger time step
$\FF\Delta t$. It is tempting to call $A$ the fast particle (or the fast state)
and $B$ the slow particle (or the slow state). This linguistic shortcut must
be used with care, because $A$ and $B$ particles share identical dynamic
properties, and this is only the way we treat them in our discretization
procedure which differs.

In practice, the exchange rates must be tailored in such a way that the
reaction zone falls into a region where particles $B$ dominate. If we are
confident that particles with state $A$ never wander around the reaction zone,
we can safely drop the terms $\exp(-\HH_{A,R}\Delta t/2)$ from the algorithm.
Moreover, there is a considerable room for optimizing the multiple time step
algorithm, at the expense of a slightly more complex programming.

If necessary, the idea can be further extended by adding more internal states
$A_1$,$A_2$, $A_3$\ldots, corresponding to consecutive nested sub-domains of
the configuration space. The corresponding generalization of the algorithm is
straightforward.

To check the relevance of the above idea, we now introduce an exchange
zone for particles diffusing in one dimensional space as described earlier.
The ``slow'' particles and ``fast'' particles exchange their status in a slab
$0.2 \leq |x|\leq 0.3$ of width $\Lambda_w=0.1$. The acceleration factor is
kept the same $\FF=8$. The average density of particles $A$ and $B$ now
overlaps (see Fig.~\ref{fig:exchange}), and the artifact shown in
Fig.~\ref{fig:sharp} seems to be removed. Thus, our
splitting~(\ref{eq:splittingMultistep}) looks like an efficient strategy for
keeping the advantages of a multiple time step algorithm, without suffering
from appreciable unphysical features.

The multiple time step algorithm shows equal efficiency in higher dimensions.
To demonstrate the advantages of the exchange zone in higher configurational
space, we consider independent particles diffusing in a $\ds$ dimensional
configuration space. Particles are reflected by a (hyper)spherical outer
boundary located at distance $r=R_e$ from the origin and become captured when
hitting an absorbing (hyper)spherical inner boundary at $r=R_i$, as depicted
in Fig.~\ref{fig:hypersphere}. This system is a straightforward
generalization of the previous unidimensional case.  Despite inherent
complexity, this reaction-diffusion situation remains amenable to an
analytical exact solution thanks to the rotational invariance of the system.
Details of the analytical solution are provided in
Appendix~\ref{sec:hyperspherical}.

For the multiple time step procedure, we use $\FF = 8 $ and set two exchange
zones of width $\Lambda_w=0.15$ at the vicinity of the inner and outer walls
(respectively $R_i=1.$ and $R_e\simeq 2.5$).  Figure~\ref{fig:exchange-D}
shows no visible artifact in the particle distribution when the exchange zone
width $\Lambda_w$ is kept finite, while the distribution is no longer uniform
if $\Lambda_w=0$ as shown in Fig.~\ref{fig:naive}.  In Fig.~\ref{fig:laplace},
we compare the numerical and analytical Laplace transforms of the survival
rate $\hat{P}_{D,R_i,R_e}(s)$ (\textit{c.f.}
Appendix~\ref{sec:hyperspherical}), starting from a uniform particle
distribution at initial time, for $\ds=3$, 5, 8 and 10. The combination of an
exchange zone and internal switching state dynamics maintains a quasi-uniform
distribution, while reducing the computation time by factor 2. This fits our
expectation as discussed in the next section.

\section{Example}
\label{sec:realisticCases}

We now turn to a realistic situation to see whether our strategy keeps its
promises. We consider, for example, two localized colloid carrying a ligand
and many receptors, respectively, on their surfaces. These colloids combine in
the form of doublet upon ligand-receptor binding. We model the rotational
diffusion of colloids as random walks of a particle on a sphere.  The motion of
a Brownian particle evolving on a sphere is parameterized in polar coordinates
$(\theta,\phi)$, with a capture zone defined by
$\theta<\theta_c$~(Fig.~\ref{fig:sphere}).  We performed a statistics of
survival times $t_s$, starting from a uniform distribution of initial
conditions over the sphere, until the particle encounters the capture circle
$\theta=\theta_c$. An analytical solution of this problem, presented
in~\cite{2008_Lee_Marques} served as a benchmark for checking out the whole
procedure.

To implement the optimization procedure, we defined an exchange zone between
angles $\theta_{inf}=0.4$ and $\theta_{sup}=0.6$, while the capture zone was
set to $\theta_c=0.03$.  Figure~\ref{fig:survival-times-sphere} shows the
survival probability as a function of time for various acceleration factors
$\FF=4$, 8, 16 and 32 without any appreciable deviation, while the angular
distribution of A and B particles is presented in
Fig.~\ref{fig:histograms-sphere-reaction-on}. Meanwhile, we must check
whether the multiple time step algorithm preserves the uniform distribution of
positions when there is no capture (reaction off), which was achieved by
counting the number of recurrences, as shown in
Fig.~\ref{fig:histograms-sphere-reaction-off}.

For a proper choice of $\FF$, we must keep the length step smaller than the
width of the exchange zone $\Lambda_w$.  Choosing a larger $\FF$ might
accelerate the fast particle but requires an accordingly larger exchange zone,
where the particle might spend a longer time. Hence, we observe that the CPU
time is reduced by factor 6 for $\FF=16$ and by 9 for $\FF=32$ with $\Lambda_w
= 0.2$ and $0.4$ respectively. Corresponding CPU times are provided in
Table~\ref{table:cpu-times}.  For a given $\FF$, the time step of fast
particle is increased by $\FF$, and the corresponding length step ($\Delta
\theta$) by $\sqrt{\FF}$.

A rough estimate of the CPU time dependence on~$\FF$ can be obtained with a
few approximations.  Provided that the volume of the interior zone (including
exchange zone) where particles are ``slow'' occupies a fraction $\alpha$ of
the total configuration space volume $L^d$, and that the remaining part of the
configuration space is filled with ``fast'' particles subject to the
acceleration factor $\FF$, then the time needed to sample the whole
configuration space is approximately proportional to
\beq
\alpha + \frac{1-\alpha}{\FF},
\label{eq:estimated-cpu-time}
\eeq
and the computing time reduced by a factor $\FF/(1-\alpha +\FF \alpha)$.  In
our case, with a the sphere of area $4\pi$ and an exchange zone $[0.4,0.6]$,
we observe a confirmation of the expected behavior (see
Fig.~\ref{fig:cpu-F}). Equation~(\ref{eq:estimated-cpu-time}) also accounts
fairly for the CPU time associated with the diffusion in $\ds$ dimension
illustrated in Fig.~\ref{fig:hypersphere} with a coefficient $\alpha$
comprised between 0.3 and 0.6.

If one now wishes to compare the CPU time of the multiple time step algorithm
with exchange procedure (Table~\ref{table:algo-I}) with its single time step
counterpart~(Table~\ref{table:algo-II}), eq.~(\ref{eq:estimated-cpu-time})
must be changed in
\beq
\bigg[\alpha + \frac{1-\alpha}{\FF}\bigg](1+\varepsilon),
\label{eq:modified-cpu-time}
\eeq
where $\varepsilon$ accounts for the relative increase of computations
associated with the exchange procedure, which depends on each specific case but
should remain small in all relevant situations.

The acceleration factor $\FF$ cannot exceed an upper value $\FF_{max}$
dictated by the exchange zone size $\Lambda_w$ and slow zone size $L
\alpha^{1/d}$. Thus, increasing significantly $\FF$ without reintroducing
spurious sampling features requires the enlargement of both exchange and slow
zones, and hence does not improve so much the efficiency.  The optimal
strategy may consist in partitioning the configuration space into nested
domains $\aA_1$, $\aA_2$, $\aA_3$\ldots associated respectively to
acceleration factors $1$, $\FF$, $\FF^2$\ldots and with geometrically
increasing characteristic sizes.  A well balanced partition should be such
that the computer time linked to crossing each domain $\aA_i$ remains
approximately constant (a procedure reminiscent from the so-called Monte-Carlo
umbrella sampling~\cite{book_Frenkel_Smit} or Transition Interface Sampling
methods~\cite{1997_Dellago_Chandler}). In practice, values of $\FF$ equal to~8
or~16 sound promising.

Finally, we considered the case of $N_L$ ligands simultaneously diffusing on a
sphere covered with $N_R$ randomly placed receptors.  There are $N_R$ capture
zones and each capture zone is defined as angular distances from each
receptor.  As discussed in ref.~\cite{2008_Lee_Marques}, one expects the
survival probability of the ligands to decay exponentially at long times, with
an inverse relaxation time proportional to the number of receptors. In
Fig.~\ref{fig:many-rtime}, we compare the survival time distributions
obtained with the single time step and multiple time step procedure.  As shown
in Fig.~\ref{fig:many-rtime}, this distribution is insensitive to the choice
of the simulation procedure with the choice of a proper $\FF$ (here $\FF =
16$).

The computational time of three different procedures is compared: single time
step algorithm, multiple time step algorithm with exchange zone and multiple
time step algorithm without exchange zone (naive procedure). The benefits of
the multiple time step algorithm decrease when the number of ligands $N_L$
increases. When a single receptor ($N_R=1$) is present on the sphere, the
multiple time step algorithm performs always better than the single time step
one (Fig.~\ref{fig:cpu-time1}). If more receptors are present ($N_R=10$), the
benefits of the multiple time step procedure with exchange zone are lost when
the number of ligands reaches $N_L=20$ (Fig.~\ref{fig:cpu-time10}).  This is
because the switching of the particle status in the exchange zone requires as
many as $N_L\times N_R$ operations, which cancels out the gain from using the
acceleration factor $\FF\times N_L$ (large $\varepsilon$ in
expression~(\ref{eq:modified-cpu-time})).  However, if pairwise interactions
among ligands (receptors) were to be considered, the additional number of
operations for particle status switching would become negligible compared with
the quadratic $N_L^2$ number of operations required by the two-body
interactions (small $\varepsilon$ in expression~(\ref{eq:modified-cpu-time})).
Thus, multiple time step algorithm is again clearly favorable.

\section{Conclusions}
\label{sec:conclusion}

We presented a systematic procedure to simulate the Brownian diffusion of
independent (non-interacting) and non conserved particles. Based on the
second-quantized formulation of Doi, our approach uses a Trotter splitting
technique and provides a weak order two algorithm for simulating the original
process.

We also showed how this reaction-diffusion approach could help in finding an
optimized and faithful sampling of the configuration space, in cases where the
spatial resolution of the Brownian paths is not necessarily homogeneous,
without showing the artifacts that other naive approaches encounter.

We tested our model on particles diffusing on a sphere and reacting at the
vicinity of a pole. We were able to cut by an order of magnitude the computing
time, while preserving the accuracy of the numerical scheme.  When many
particles diffuse simultaneously in the presence of multiple reacting zones,
the efficiency of the multiple time step procedure declines for large number
of particles.  Although the efficiency depends on the overall algorithmic
complexity of the various parts of the simulation scheme, the computation time
of multiple time step algorithm is bounded by the computation time of single
time step algorithm.

In the multiple time step algorithm, our trick consists in introducing an
extra state variable which stochastically switches between two internal
states. This procedure is certainly not restricted to the Brownian dynamics
exemplified above, but might be also of interest for deterministic, or
inertial Langevin dynamics. This formalism will be the natural starting point
for such generalization.

\textbf{Acknowledgement}
Author F.T. thanks KIAS for financial support during its visit.
N.-K. L. acknowledges financial support by the Korea Research Foundation Grant
funded by the Korean Government (MOEHRD, Basic Research Promotion Fund)
(KRF-2005-204-C000024).

\clearpage
\appendix
\section{Connection between second-quantized formulation and master equation}
\label{sec:connection}

Starting from
\beq
n(\br,t) = \bra{0}a(\br)e^{-\HH\, t}\ket{\Phi(0)}
\label{eq:evolution}
\eeq
one splits the evolution operator
\beq
e^{-\HH\, t} = e^{-\HH\, (t-t')}\cdot e^{-\HH\, t'}
\eeq
and introduce a projector (valid only in the subspace of states with no
multiple occupancy)
\beq
\bm{1} = \bigg( \ket{0}\bra{0} + \int\dd\br
\ad(\br)\ket{0}\bra{0}a(\br)\bigg),
\label{eq:unity}
\eeq
leading  to
\beq
n(\br,t) = \int\dd\br'\bra{0}a(\br)e^{-\HH (t-t')}\ad(\br')\ket{0}
\bra{0}a(\br')e^{-\HH t'}\ket{\Phi(0)}
\eeq
and to the well known propagator structure of the time evolution~:
\beq
n(\br,t) = \int\dd\br'G(\br,\br';t-t')n(\br',t')
\eeq

In the same way, the partial differential
equation~(\ref{eq:standardMasterEquation}) is obtained by differentiating
eq.~(\ref{eq:evolution})~:
\beqn
\partial_t n(\br,t)
&=& -\bra{0}a(\br)\HH e^{-\HH t}\ket{\Phi(0)}\notag\\
&=& -\bra{0}\HH a(\br)e^{-\HH t}\ket{\Phi(0)}
+\bra{0}[\HH,a(\br)]e^{-\HH t}\ket{\Phi(0)}.
\label{eq:toReactionDiffusionEquation}
\eeqn
The Hamiltonian is a sum of normal ordered (creator operators on the left of
annihilator operators) terms and pure annihilator terms $a(\br)$, which ensures
that only the second term of~(\ref{eq:toReactionDiffusionEquation})
is non zero. A standard commutator calculation leads to
eq.~(\ref{eq:standardMasterEquation}).

The connection between the second-quantized formulation of the dynamics and
the algorithm is done by splitting the evolution operator, such as
in~(\ref{eq:splittingRFDDFR}) or in eq.~(\ref{eq:splittingMultistep}), and by
introducing as many projection operators~(\ref{eq:unity}) as necessary between
the exponentials. Each one of the $\bra{0}a(\br_1) e^{-\HH \Delta t}
\ad(\br_2)\ket{0}$ has a straightforward interpretation, and corresponds to an
elementary step of our algorithms~I and~II.

\section{Heun and Collatz schemes for a first order differential equation}
\label{sec:HeunCollatz}
For integrating an ordinary differential equation (o.d.e) $\dot{\br}=\bf(\br)$,
the simplest numerical scheme is the Euler method~:
\beq
\br(t+\Delta t) = \br(t)+\Delta t \bf(\br(t)).
\label{eq:EulerScheme}
\eeq
However, this scheme is known to be an ``order one'' integrator, which in the
present situation, means that the approximated solution departs from the exact
one, on a time interval $\Delta t$, by an amount of order $C \Delta
t^2$. Clearly, this error is not acceptable as it would spoil the accuracy of
the whole algorithms~I and~II.

Among the number of existing ``order two'' integrator, we propose to use
either~:
\beq
\br(t+\Delta t) = \br(t)+\frac{\Delta t}{2}
\bigg[\bf(\br(t)) + \bf\bigg(\br(t)+\Delta t\bf(\br(t))\bigg)\bigg],
\label{eq:HeunScheme}
\eeq
(Heun scheme), or
\beq
\br (t+\Delta t) = \br(t)+ \Delta t
\bigg[\bf\bigg(\br(t)+\frac{\Delta t}{2}\bf(\br(t))\bigg)\bigg],
\label{eq:CollatzScheme}
\eeq
(Collatz scheme)~\cite{book_Stoer_Bulirsch}. These schemes both require two
evaluations of the function $\bf(\br)$ each, per time step. Once $\Delta t$
has been chosen, it is possible to give to eq.~(\ref{eq:HeunScheme})
and~(\ref{eq:CollatzScheme}) a form similar to~eq.~(\ref{eq:EulerScheme}), by
defining a pseudo force field~:
\beq
\btf(\br) = \frac{1}{2}
\bigg[\bf(\br) + \bf\bigg(\br + \bf(\br) \Delta t\bigg)\bigg]
\label{eq:HeunPseudoforce}
\eeq
(Heun pseudo force field), and~:
\beq
\btf(\br) = \bf\bigg(\br + \bf(\br)\frac{\Delta t}{2}\bigg)
\label{eq:CollatzPseudoforce}
\eeq
(Collatz pseudo force field),
such that~:
\beq
\br(t+\Delta t) = \br(t)+\Delta t\, \btf(\br(t)).
\label{eq:OrderTwoScheme}
\eeq
becomes a genuine order two integrator. We suggest to use either
(\ref{eq:CollatzPseudoforce}) or (\ref{eq:OrderTwoScheme}) in the
steps of the algorithms associated to $e^{-\HH_F\Delta t}$.

\section{Diffusion and absorption in $\ds$ dimensions}
\label{sec:hyperspherical}

To compute the survival probability between an absorbing boundary at $r=R_i$
and a reflecting boundary at $r=R_e$, we introduce the concentration of
particle $c(r,t)$ at position $r$ and time $t$, obeying: 
\beq
(\partial_t  -\Delta) c = 0,
\eeq
with $c(R_i,t)=0$,  $\partial_r c(R_e,t)=0$ and where 
\beq
\Delta = \frac{\ds-1}{r}\frac{\dd}{\dd r} + \frac{\dd^2}{\dd r^2} 
\eeq
is the hyperspherical Laplacian in $\ds$ dimensions, discarding everything but
the radial variable $r$. In this example, the diffusion constant $D$ is set to
one without loss of generality.  Denoting $c_0=c(r,0)$ the initial uniform
concentration, the time Laplace transformed $\hat{c}(r,s)$ obeys
\beqn 
\Big(s - \frac{\ds-1}{r}\frac{\dd}{\dd r} - \frac{\dd^2}{\dd r^2}\Big)
\hat{c}(r,s) &=& c_0;\\
 \frac{2\pi^{\ds/2}}{\Gamma(\ds/2)}\int_{R_i}^{R_e} \dd r\,
 r^{\ds-1}\, c_0 &=& 1, 
 \eeqn
the solution of which can be expressed in terms of modified Bessel functions,
with $\nu=\ds/2-1$:
\beq
\hat{c}(r,s) = \frac{c_0}{s}\Bigg[1 - \frac{R_i^{\nu}}{r^{\nu}}%
\frac{ I_{\nu+1}(\sqrt{s}R_e) K_{\nu}(\sqrt{s}\,r)+
  K_{\nu+1}(\sqrt{s}R_e)I_{\nu}(\sqrt{s}\,r)}
{I_{\nu+1}(\sqrt{s}R_e)K_{\nu}(\sqrt{s}R_i)
 +K_{\nu+1}(\sqrt{s}R_e)I_{\nu}(\sqrt{s}R_i)}\Bigg]. 
\eeq
Integrating $\hat c(r,s)$ over $r$ between $R_i$ and $R_e$ gives the Laplace
transform $\hat{P}_{\ds,R_i,R_e}(s)$ of the survival rate $P_{\ds,R_i,R_e}(t)$:
\beq
\hat{P}_{\ds,R_i,R_e} (s)=
\frac{1}{s}\Bigg[1 +\frac{\ds}{\sqrt{s}R_i}
\frac{1}{\bigg[(\frac{R_e}{R_i})^\ds-1\bigg]}  
\frac{I_{\dsd}(\sqrt{s}R_i)K_{\dsd}(\sqrt{s}R_e)
 -I_{\dsd}(\sqrt{s}R_e)K_{\dsd}(\sqrt{s}R_i)}
{I_{\dsd}(\sqrt{s}R_e)K_{\dsd-1}(\sqrt{s}R_i)
 +K_{\dsd}(\sqrt{s}R_e)I_{\dsd-1}(\sqrt{s}R_i)}
\Bigg].
\label{eq:ExactLaplace}
\eeq
The survival rate $P_{\ds,R_i,R_e}(t)$ is a monotonically decreasing function
of time $t$, giving the proportion of particles remaining after $t$, starting
from a uniformly distributed random initial position in configuration space.
It is related to the average survival time $\mean{t}$ by the simple relations:
\beq
\mean{t} = \int_O^{\infty} \dd t\, P_{\ds,R_i,R_e}(t) = \lim_{s\to 0}
\hat{P}_{\ds,R_i,R_e}(s). 
\eeq

The simulation provides us with a finite sample of $N$ survival times $\{t_i\}$,
$1\leq i\leq N$. These survival times can be reordered into a monotonically
increasing sequence $t_{\sigma(i)}$ by means of a suitable permutation
$\sigma$ of the indexes. Such a numerical sample approximates the survival
rate $P_{\ds,R_i,R_e}(t)$ up to statistical fluctuations of order
$1/\sqrt{N}$. 

In practice, this is achieved by plotting $1-i/N$ as a function of
$t_{\sigma(i)}$.  Alternatively this sample can be used to numerically obtain
the Laplace transform of the survival rate $\hat{P}_{\ds,R_i,R_e}(s)$, by
plotting $\hat{P}^{(\mathrm{num})}(s)$ versus $s$, as shown in
Fig.~\ref{fig:laplace}, with $\hat{P}^{(\mathrm{num})}(s)$ defined as
\beq
\hat{P}^{(\mathrm{num})}(s) = \frac{1}{N}\sum_{i=1}^{N}
\frac{1-\exp(-s\,t_{\sigma(i)})}{s}.
\label{eq:NumericalLaplace}
\eeq

\clearpage
\section*{Figures and Tables}

\begin{table}[b]
\begin{center}

\begin{tabular}{l}
\hline\hline
\textbf{Algorithm~I: reaction-diffusion over} $\Delta t$\\
\hline
\textbf{1.}~Consider the probability that a particle located at $\br(t)$ reacts
during a time interval $\Delta t/2$,\\
given that the probability of this event is \hbox{$1-\exp(-S(\br(t))\Delta
  t/2)$}.  If the particle reacts, end the\\
procedure and set the reaction time $t_R$ to $t+\Delta t/2$.\\
\textbf{2.}~Shift the position of the particle by $\mu\btf(\br(t))\Delta t/2$,
as a consequence of the drift force. \\
\textbf{3.}~Draw from a Gaussian distribution of variance $2 D\Delta t$ a
random step $\Delta \br^D$, and shift the\\
position by $\Delta \br^D$.\\
\textbf{4.}~Shift the position by $\mu\btf\Big(\br(t) +\mu\btf(\br(t))\Delta
t/2 +\Delta \br^D\Big)\Delta t/2$. The particle sits now at a\\
position $\br(t+\Delta t)$\\
\textbf{5.}~Consider the probability that a particle located at $\br(t+\Delta
t)$  reacts during a time interval\\
$\Delta t/2$. As in step~1, if the reaction takes place, set the reaction time
to $t_R=t+\Delta t/2$\\
\hline\hline
\end{tabular}

\caption{Algorithm for a coordinates dependent $X\to 0$ process}
\label{table:algo-I}

\end{center}
\end{table}

\clearpage
\begin{table}
\begin{center}

\begin{tabular}{l}
\hline\hline
\textbf{Algorithm~II: multiple time step algorithm}\\
\hline
\textbf{1.}~Repeat $\FF/2$ times\\
\hspace{0.3cm}\textbf{1.a.}~If the particle is $B$, consider changing its state
during  a time interval $\Delta t/2$, else do nothing.\\
\hspace{0.3cm}\textbf{1.b.}~If the particle is still $B$, proceed with
algorithm~I  during the time interval $\Delta t$.\\
\hspace{0.3cm}\textbf{1.c.}~ If the particle is still $B$, consider changing
its state during a time interval $\Delta t/2$, else do nothing.\\
\textbf{2.}~Do once\\
\hspace{0.3cm}\textbf{2.a.}~If the particle is $A$, consider changing its
state during  a time interval $\FF\Delta t/2$, else do nothing.\\
\hspace{0.3cm}\textbf{2.b.}~If the particle is still $A$, proceed with
algorithm~I during the time interval $\FF\Delta t$.\\
\hspace{0.3cm}\textbf{2.c.}~If the particle is still $A$, consider changing
its state during a time interval $\FF\Delta t/2$, else do nothing.\\
\textbf{3.}~Repeat $\FF/2$ times\\
\hspace{0.3cm}\textbf{1.a.}~If the particle is $B$, consider changing its
state during a time interval $\Delta t/2$, else do nothing.\\
\hspace{0.3cm}\textbf{1.b.}~If the particle is still $B$, proceed with
algorithm~I during the time interval $\Delta t$.\\
\hspace{0.3cm}\textbf{1.c.}~ If the particle is still $B$, consider changing
its state during a time interval $\Delta t/2$, else do nothing.\\
\hline\hline
\end{tabular}

\caption{Multiple time step algorithm built on algorithm~I}
\label{table:algo-II}

\end{center}
\end{table}

\clearpage
\begin{table}
\begin{center}
\begin{tabular}{|c|c|c|c|c|}
\hline\hline
factor $\FF$ & time (s) & zone $[\theta_{inf},\theta_{sup}]$ & $\Lambda_w$ &
CPU time ratio\\
\hline
1 & 299.8 & $--$ &  $--$ &1.0\\
\hline
4 & 106.0 & [0.4-0.6] & 0.2 & 0.354 \\
\hline
8 & 62.0 & [0.4-0.6] & 0.2 & 0.207\\
\hline
12 & 51.2 & [0.4-0.6] & 0.2 & 0.171\\
\hline
16 & 42.6 & [0.4-0.6] & 0.2 & 0.142\\
\hline
16 & 45.1 & [0.3-0.7] & 0.4 & 0.150\\
\hline
32 & 33.5 & [0.3-0.7] & 0.4 & 0.112\\
\hline\hline
\end{tabular}
\end{center}
\caption{Summary of CPU times relative to the absorption of 1000 independent
  particles at $\theta_c=0.03$. Brackets indicate the limits of the exchange
  zone. See text for details.}
\label{table:cpu-times}
\end{table}

\clearpage

FIGURE CAPTIONS\\

FIGURE~1.\\
This picture shows the principle of a multiple time step procedure for a
reaction-diffusion process. Finding the small reaction well requires the
step size between two consecutive spatial positions to be smaller than the
size of the well. Large steps might miss the reaction well completely, and the
simulated reaction rate can only be largely underestimated. However, far from
the reaction zone, large steps cause no harm, and can speed up the exploration
of the configuration space by a large factor. Nevertheless, the boundary
between the two zone (dashed circular line) will be subject to sampling
artifacts if the procedure is done without care.
\medskip

FIGURE~2.\\
This graph presents the spurious features expected from a naive multiple time
step procedure in the absence of exchange zone. When an accelerating factor
$\FF=8$ or $\FF=32$ is introduced, two peaks appear around $x=\pm 0.2$. Their
width directly depend on the length step, namely $\sqrt{2 D\FF\Delta t}$.
This artifact comes from a local breakdown of detailed balance induced by the
time discretization. The special case $\FF=1$, with no acceleration, gives a
control curve, constant as expected, up to statistical fluctuations.  
\medskip
  
FIGURE~3.\\
Particle density in the absence of reaction when dynamics is governed by the
internal state exchange dynamics in one dimension. ``Slow'' $B$ particles
occupy the central regions, and ``fast'' $A$ particles mainly occupy the
outside of the switching zone. The flat curves show the joined $A$ and $B$
distribution, which remains constant. When the exchange rates are increased,
the $A/B$ proportion tends towards its limit value~(\protect
\ref{eq:molarFractions}), here a piecewise linear function. The exchange
region where rates $w_{AB}(\br)$ and $w_{BA}(\br)$ are both non zero is the
interval $[0.2,0.3]$ corresponding to an exchange zone size $\Lambda_w=0.1$.
The acceleration factor used in this case is $\mathcal{F}=8$. The artifact in
Fig.~\protect\ref{fig:sharp} is no longer visible.
\medskip
\clearpage

FIGURE~4.\\
Schematic view of the $\ds$-dimensional configuration space, bounded by a
reflecting outer boundary $R_o$ and an absorbing inner boundary $R_i$. Two
exchange zones are indicated by grey areas limited by dashed lines.

\medskip

FIGURE~5.\\
Repartition (density) of particles as a function of $r$, with $R_e\simeq 2.5$,
$R_i=1.0$ and $\ds=5$. The exchange zone size $\Lambda_w$ here approaches
0.15. The inset shows an enlargement of the inner exchange zone.  \medskip

FIGURE~6.\\
When the exchange zone is shrunk to $\Lambda_w=0$, under the same conditions
as ~Fig.~\protect\ref{fig:exchange-D}, the unphysical heterogeneity
(artefact) appears near the outer exchange zone ($r\sim 2.3$) and, to a lesser
extent, near the inner exchange zone ($r\sim 1.15$ as shown on the inset of
this graph).  \medskip

FIGURE~7.\\
Comparison of the survival rate, in Laplace space, between simulations and
exact result, for spatial dimensions $\ds=3, 5, 8$ and $10$ (see \textit{e.g.}
equations~(\ref{eq:ExactLaplace}) and~(\ref{eq:NumericalLaplace}) in
Appendix~\ref{sec:hyperspherical}).  We use for the simulations an
acceleration factor $\mathcal{F}=8$ and a sample size of $N=1000$ independent
particles. The horizontal axis is the Laplace variable $s$ while the vertical
axis has the dimension of a time.  The extrapolated value as $s \to 0$ gives
the average residence time $\mean{t}$ of the particles, and is subject to a
statistical uncertainty of order $\mean{t}/\sqrt{N}$ (the distribution of
residence times is nearly exponential).  \medskip

FIGURE~8.\\
Pictorial view of the reaction-diffusion problem on a sphere. The exchange
zone $A \leftrightarrow B$ is indicated by the shaded area. Near the capture
zone, $B$ particles are subject to the small step dynamics while apart from
it, $A$ particles are subject to the large step dynamics.  
\medskip

FIGURE~9.\\
The mean survival time $\mean{\tau_r}$ is calculated as a function of reaction
zone size $\theta_c$ using different values of the acceleration~$\FF$ (4 to 32
indicated as symbols).  The simulation results show perfect agreements with
analytic results: $\mean{\tau_r} =-\ln(1-\cos(\theta_c))-1+\ln 2
+(1-\cos(\theta_c))/2 $, regardless of different values of $\FF$.  The time is
measured in the unit of rotational diffusion coefficient
$D_r$~\cite{2008_Lee_Marques}.  The exchange zone $\Lambda_w=0.4$ is chosen to
be $(\theta_c+0.15,\theta_c+0.55)$.
\medskip
\clearpage

FIGURE~10.\\
Recurrence histogram of A and B particles as a function of angular positions
$\theta$. Particles disappear as soon as they encounter the capture radius
$\theta_c=0.03$ and an exchange zone is included in the region $0.4\leq \theta
\leq 0.6$.  Solid lines correspond to A particle with $\FF=8, 16, 32$ from top
to bottom.  Symbols are for B particles with the same acceleration factors. A
stationary profile is expected to be reached when particles are injected as
the same rate as they are captured. The inset displays possible artifact
(kinks) near the exchange zone caused by too large $\FF$ values, with fast
particles~A able to diffuse throughout the exchange zone and to reach
$\theta_c$. Nevertheless, the capture rate and the survival time distribution
show little dependence on the choice of $\FF$
(\textit{cf}~Fig.~\protect\ref{fig:survival-times-sphere}).  \medskip

FIGURE~11.\\
Recurrence histogram of A and B particles in the interval of [$\theta$,
$\theta+d\theta$] in the absence of reaction. The exchange zone is located in
the interval $0.4\leq \theta \leq 0.6$.  Bin counts are normalized by
$\sin(\theta)$, so that uniform particle density on a sphere gives a flat
histogram.  The relative proportion of $A$ and $B$ particles changes smoothly
across the exchange zone whereas the joined histogram remains flat. Examples
are computed for $\FF=16$ and $\FF=32$.  The total number of particles in the
case of $\FF=16$ is twice as large as that of $\FF=32$, for given total
exploration time.
\medskip

FIGURE~12.\\
CPU time \textit{vs} $1/\FF$ for values of $\FF$ ranging between 4 and 256. It
confirms the trend CPU time $\sim\alpha + (1-\alpha)/\FF$ with a fraction
$\alpha\simeq 0.06$ ($\theta\simeq 0.5$), consistent with the exchange zone
$[0.4,0.6]$. Note that large values of $\FF$ are not only time-inefficient,
but also reintroduce the unwanted spurious features, calling for an optimal
compromise.  
\medskip

FIGURE~13.\\
Reaction times are evaluated by three different numerical schemes: ($\star$) a
multiple-time-step procedure ($\FF=16$) without any exchange zone (naive
approach), ($\circ$) a multiple-time-step procedure ($\FF=16$) with 
exchange zone and single time step procedure $\FF=1$ (dashed line).  For the
given number of receptors ($N_R=1,5,10$), reaction times are measured as a
function of the number of ligands $N_L$.
\medskip

FIGURE~14.\\
Comparison of the computation times using three algorithms.  In the presence
of a single receptor $N_R=1$, the computation times for 500 reactions are
measured as a function of the number of ligands $N_L$.
\medskip

FIGURE~15.\\
Comparison of computational times using three algorithms. For given
$N_R=10$, the computation time required for 500 reactions are measured as a
function of $N_L$. See text for details.

\clearpage
\clearpage
\begin{figure}
\begin{center}
\resizebox{\textwidth}{!}{\includegraphics*{620848JCP2.eps}}
\caption{}
\label{fig:sharp}
\end{center}
\end{figure}

\clearpage
\begin{figure}
\begin{center}
\resizebox{\textwidth}{!}{\includegraphics*{620848JCP3.eps}}
\caption{}
\label{fig:exchange}
\end{center}
\end{figure}

\clearpage
\begin{figure}
\begin{center}
\resizebox{\textwidth}{!}{\includegraphics*{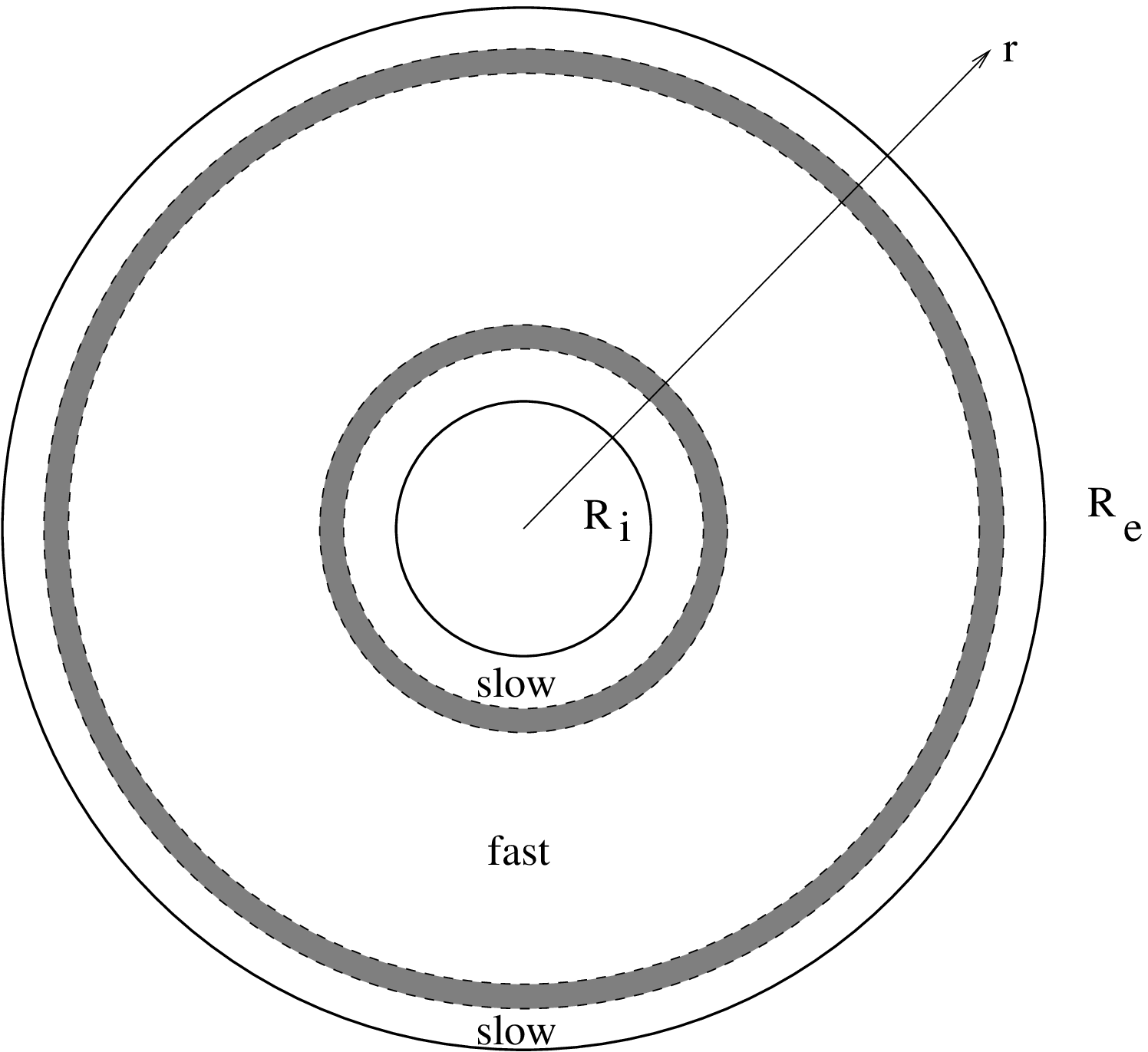}}
\caption{}
\label{fig:hypersphere}
\end{center}
\end{figure}
%

\clearpage
\begin{figure}
\begin{center}
\resizebox{\textwidth}{!}{\includegraphics*{620848JCP5.eps}} 
\caption{}
\label{fig:exchange-D}
\end{center}
\end{figure}
%

\clearpage
\begin{figure}
\begin{center}
\resizebox{\textwidth}{!}{\includegraphics*{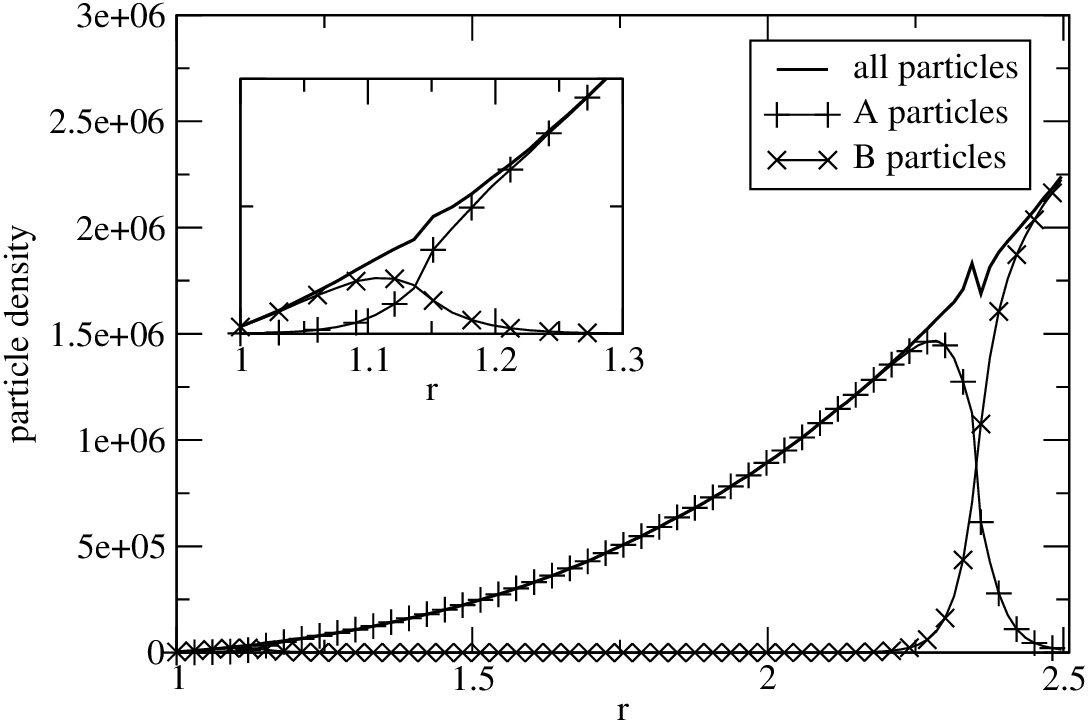}}
\caption{}
\label{fig:naive}
\end{center}
\end{figure}
%

\clearpage
\begin{figure}
\begin{center}
\resizebox{\textwidth}{!}{\includegraphics*{620848JCP7.eps}}
\caption{} 
\label{fig:laplace}
\end{center}
\end{figure}

\clearpage
\begin{figure}
\begin{center}
\resizebox{\textwidth}{!}{\includegraphics*{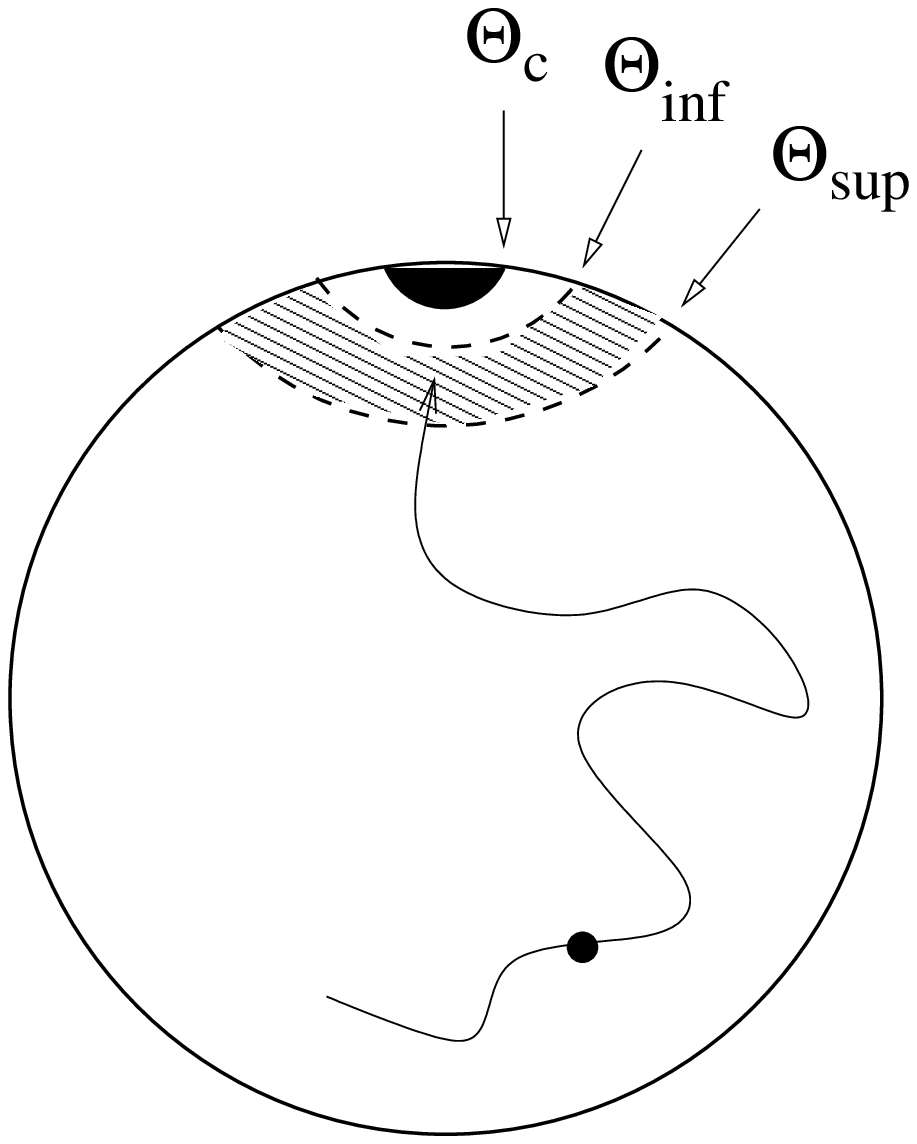}}
\caption{}
\label{fig:sphere}
\end{center}
\end{figure}
%

\begin{figure}
\begin{center}
\resizebox{\textwidth}{!}{\includegraphics*{620848JCP9.eps}}
\caption{}
\label{fig:survival-times-sphere}
\end{center}
\end{figure}
\clearpage

\begin{figure}
\begin{center}
\resizebox{\textwidth}{!}{\includegraphics*{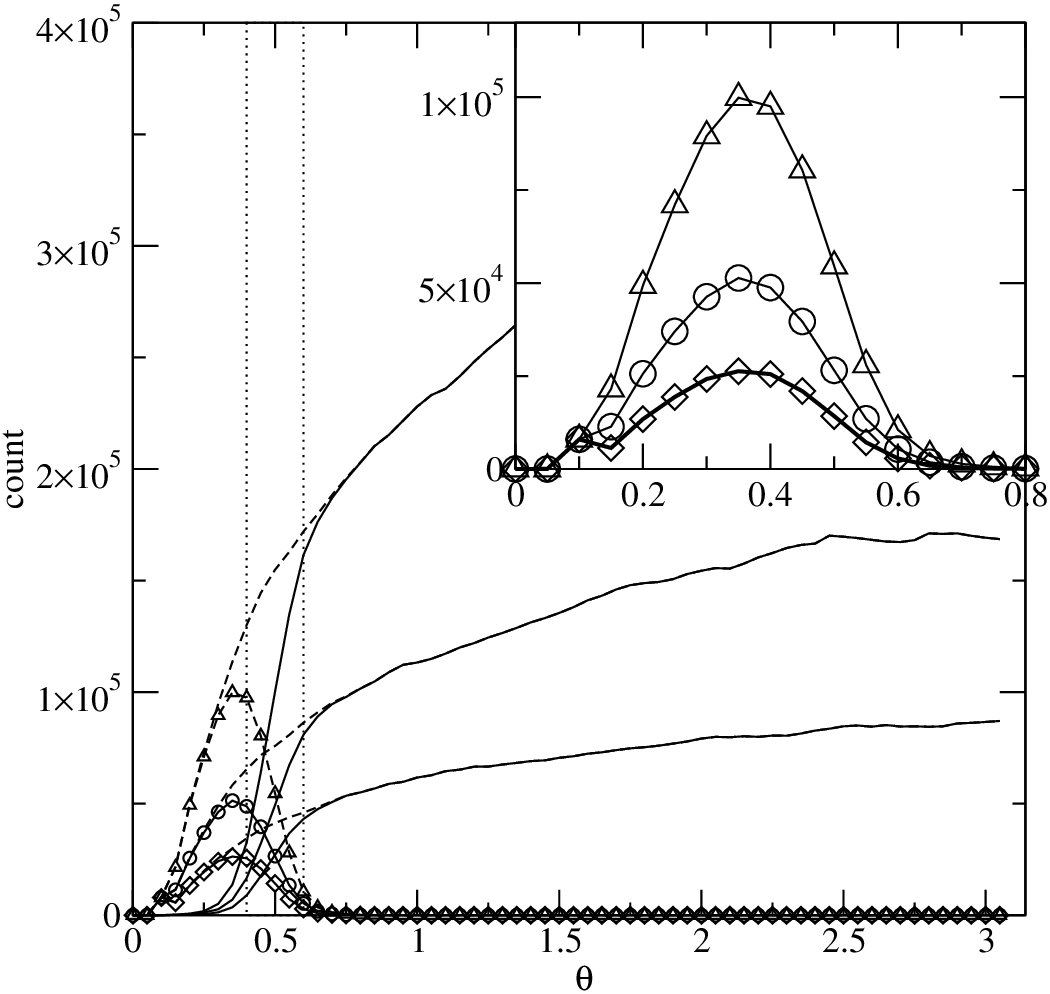}}
\caption{}
\label{fig:histograms-sphere-reaction-on}
\end{center}
\end{figure}

\clearpage
\begin{figure}
\begin{center}
\resizebox{\textwidth}{!}{\includegraphics*{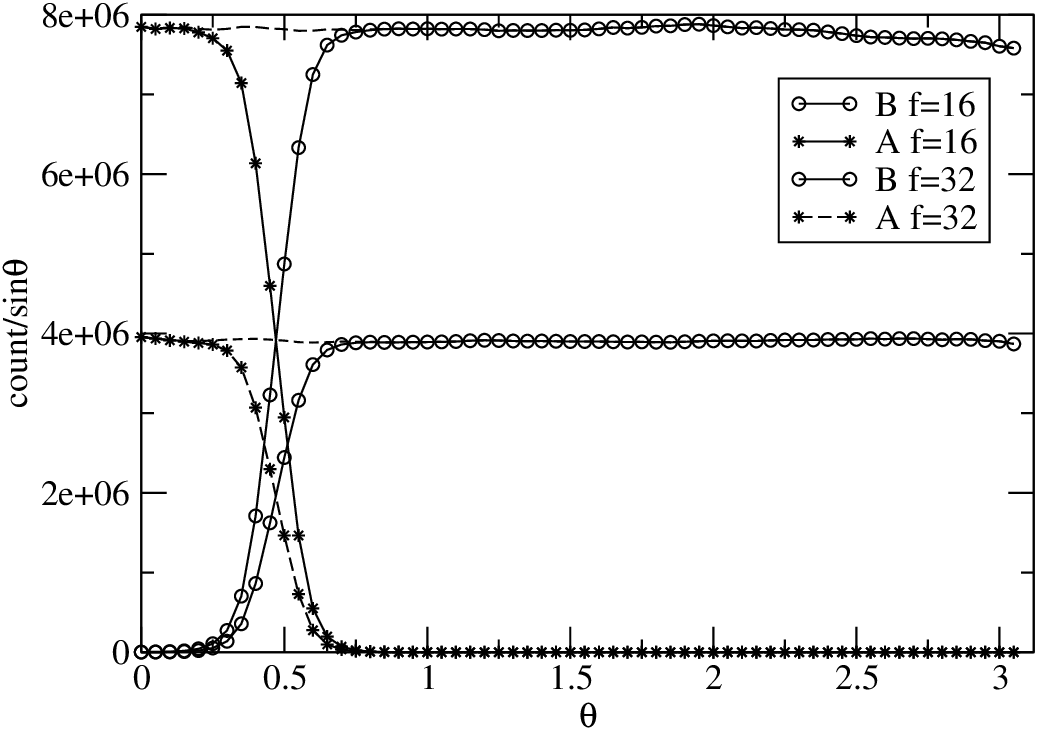}}
\caption{}
\label{fig:histograms-sphere-reaction-off}
\end{center}
\end{figure}

\clearpage
\begin{figure}
\begin{center}
\resizebox{\textwidth}{!}{\includegraphics*{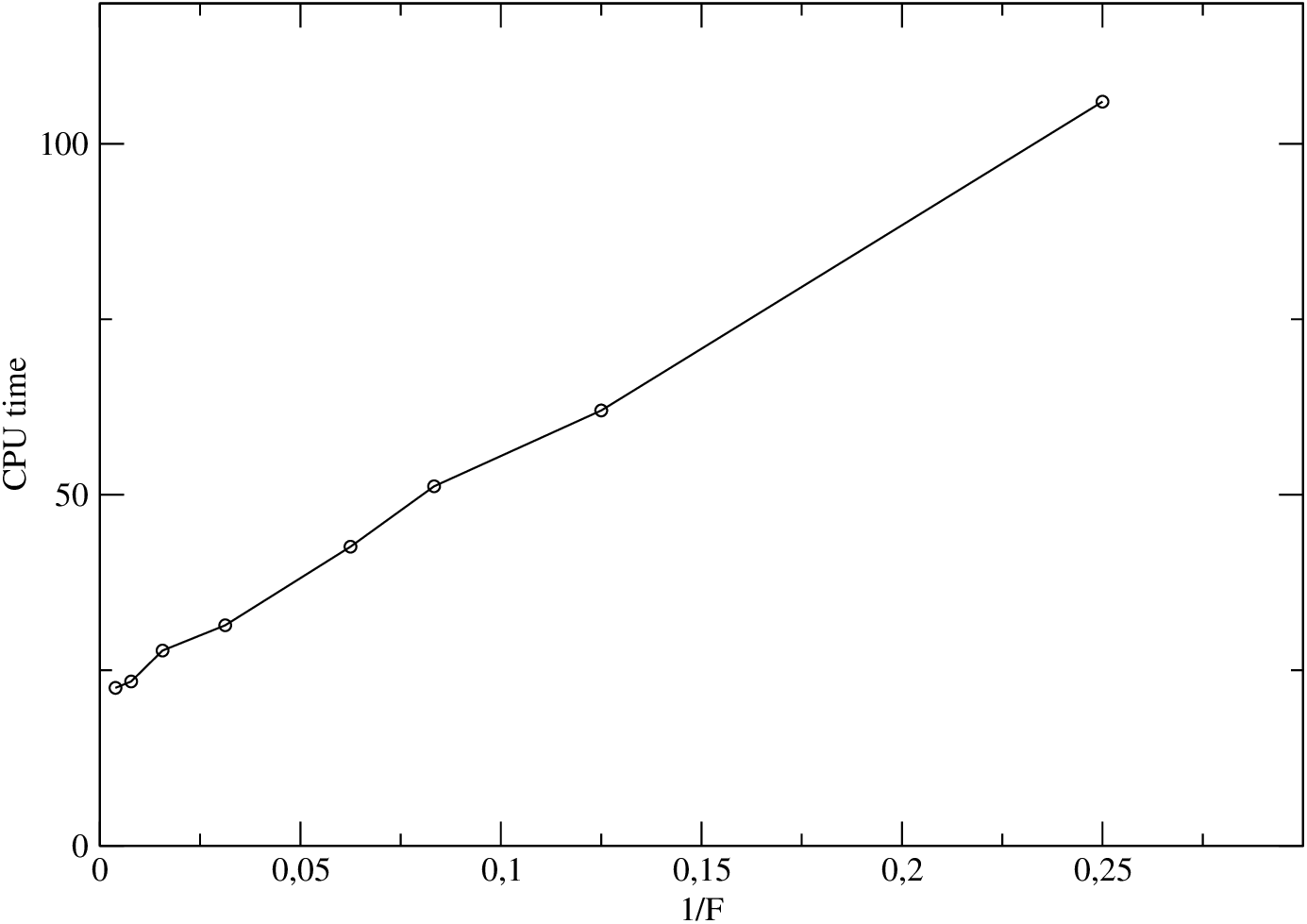}}
\caption{}
\label{fig:cpu-F}
\end{center}
\end{figure}

\clearpage
\begin{figure}
\begin{center}
\resizebox{\textwidth}{!}{\includegraphics*{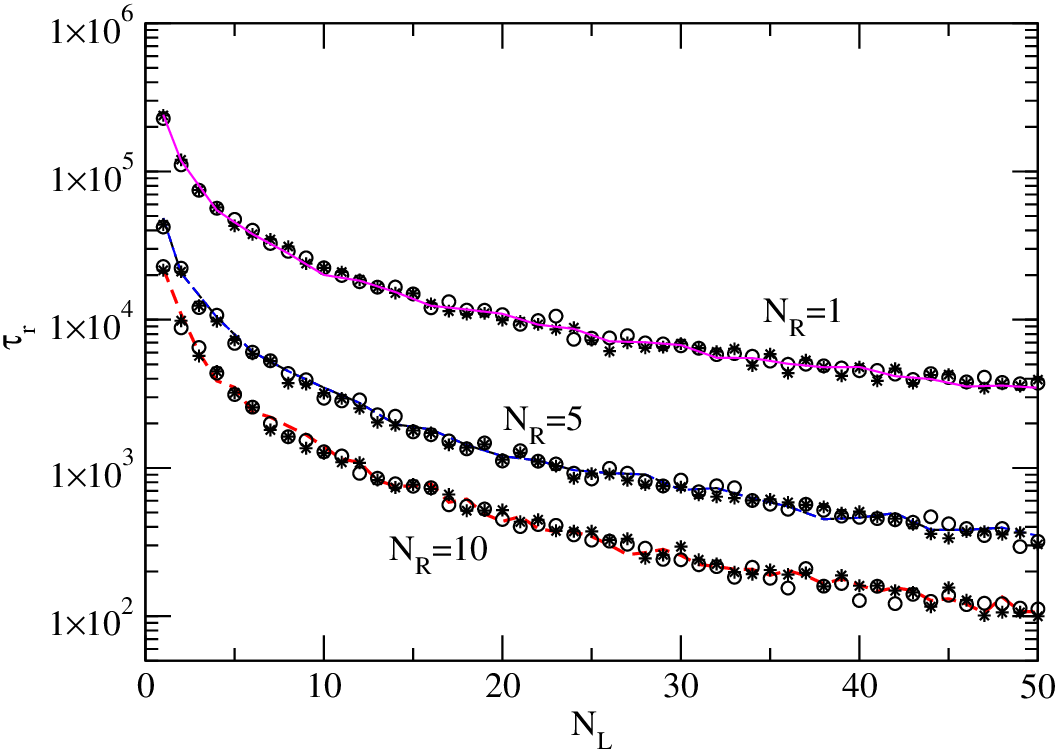}}
\caption{}
\label{fig:many-rtime}
\end{center}
\end{figure}

\clearpage
\begin{figure}
\begin{center}
\resizebox{\textwidth}{!}{\includegraphics*{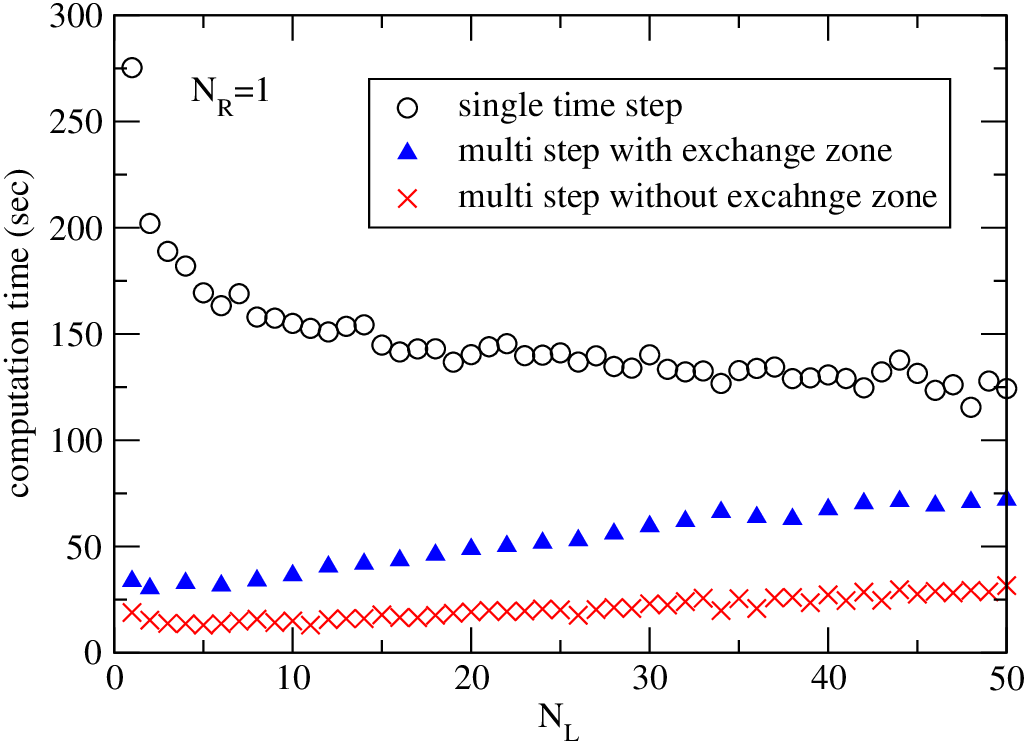}}
\caption{}
\label{fig:cpu-time1}
\end{center}
\end{figure}

\clearpage
\begin{figure}
\begin{center}
\resizebox{\textwidth}{!}{\includegraphics*{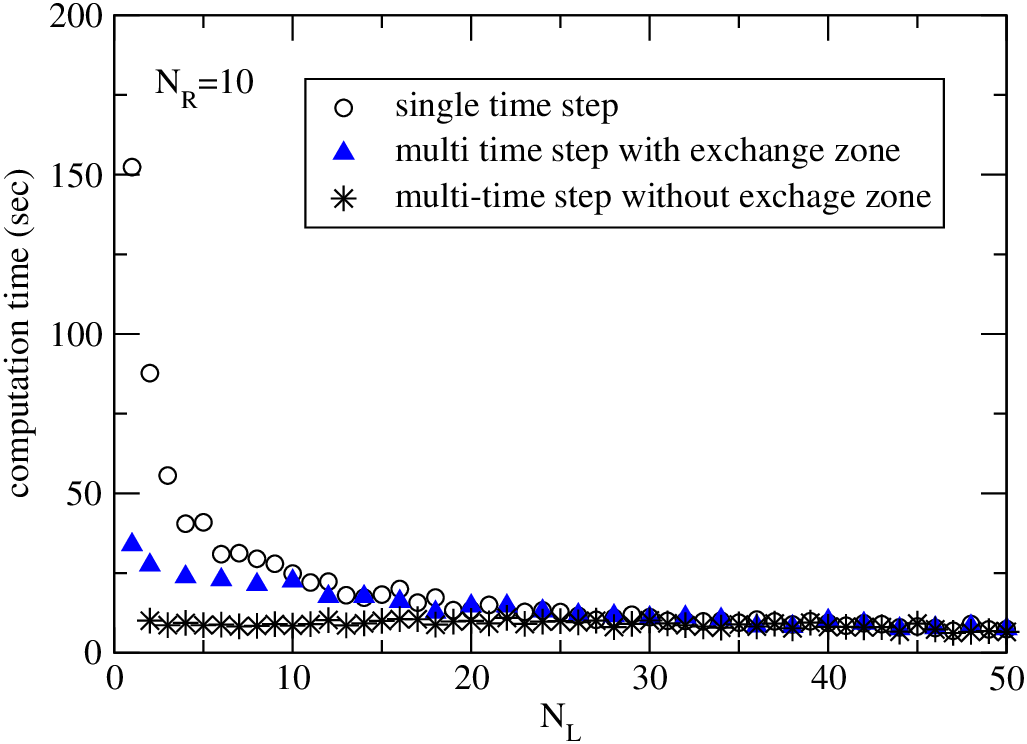}}
\caption{}
\label{fig:cpu-time10}
\end{center}
\end{figure}

\clearpage

\end{document}